\documentstyle[prb,aps,multicol,epsfig]{revtex}

\begin{document}

\title{Symmetry Considerations for detection of Time-reversal
Breaking Phases in Cuprates by X-ray Diffraction and Absorption.}
\author{S. Di Matteo$^a$ and C.M. Varma$^b$}
\address{$^a$INFM, UdR Roma III, and INFN, Laboratori Nazionali di Frascati, cp 13, I-00044   \\
$^b$Bell Laboratories, Lucent Technologies, Murray Hill, NJ 07974 }

\date{\today}

\maketitle

\begin{abstract}

Time-reversal and inversion breaking in the pseudogap region of the 
cuprate compounds have been proposed and detected by means of ARPES in the compound 
Bi-2212.
Given the implication of the effect in the physics of the cuprate superconductors,
 it is important to check the conclusions through quite 
independent experimental techniques.
 We show how  time-reversal and inversion breaking may be revealed in the Cuprates 
through resonant x-ray scattering and through non-reciprocal or magnetochiral dichroism.

\end{abstract}

\begin{multicols}{2}

 \section{Introduction}
 Time-reversal breaking Phases are predicted to occur in the so
 called {\it pseudogap} region of the phase diagram of the
 cuprates.\cite{cmv1} These phases do not change the translational symmetry
 of the lattice and therefore do not produce new Bragg spots in
 ordinary diffraction experiments. Since the spontaneous magnetic field
generated averages to zero in each unit cell, no simple magnetic dichroism effects 
arise. However,
 it was shown that Time-reversal breaking phases produce angle
 dependent dichroism in photoemission with circularly polarized
 photons.\cite{cmv2,simon} One such experiment has indeed detected such a phase.\cite{kaminski} In
 this phase both time-reversal symmetry and inversion are broken
 but their product is preserved. (Such a symmetry is called {\it
 Magnetoelectric}.\cite{LL,birss}) If this experiment is correct, the principal 
features of the theory of cuprates may be said to have been found. It is therefore
important to have independent experiments of quite a different kind
 verifying the existence of such phases. Here we propose some
 experimental possibilities, using x-ray resonant diffraction and absorption 
techniques. Another possibility using second harmonic generation has also been suggested.
\cite{mes}

 Magnetoelectric phases are known to produce non-reciprocal dichroism in
 resonant x-ray absorption in certain specified geometries. We
 present here the symmetry considerations to find the geometries
 for specific crystals of the cuprate family.
 In certain crystals (body-centered tetragonal or orthorhombic,
 for example) certain Bragg spots are forbidden due to
 interference between scattering from different atoms in the unit
 cell. Were such samples to have a transition to magnetoelectric
 phases, such Bragg spots would be allowed for some pattern of currents 
 (which enlarge the {\it primitive} unit cell) due to transition
 amplitude tensors for x-ray scattering which also break time-reversal and
 inversion. Even in the case that the current pattern does not enlarge the primitive unit cell, it is possible to detect such a phase, either by means of x-ray non-reciprocal or magnetochiral dichroism, or by means of a polarization analysis of the scattering signal. In section II we present an analysis for this processes valid for the x-ray scattering. In section III we show what can be done with the different kind of dichroisms and finally, in section IV, consider some specific crystal symmetries in the cuprate family and illustrate the geometrical conditions to show the effect on these materials.

As the effect of time-reversal and inversion breaking is purely orbital,
 we shall deal in the following with resonant scattering and absorption at the K-edge,
 where only orbital quantities are involved.\cite{carra1,luo} The generalization 
to any other edge is straightforward.

\section{Time-reversal breaking processes in x-ray resonant scattering amplitudes.}

X-ray elastic resonant scattering has become a well-known technique to
 detect magnetic signals in single crystals.\cite{luo,trammel}

In the resonant regime, the structure factor for anomalous diffraction is given by 

\begin{equation}
F({\bf q},{\bf k},\omega) = \sum_j S_j({\bf q},{\bf k},\omega) e^{i{\bf q}\cdot {\bf R_j}} 
\label{strucfac}
\end{equation}

\noindent where ${\bf k}$, ${\bf k'}$ are the wave vectors of the incoming and outgoing photon, respectively, ${\bf{k}}-{\bf{k'}}={\bf{q}}$  and $e^{i{\bf q}\cdot {\bf R_j}}$ is the usual Bragg factor. $S_j({\bf q},{\bf k},\omega)$ is the energy-dependent atomic scattering amplitude for the atom at site $j$. It can be obtained by expanding the Hamiltonian for electrons in an electromangetic field up to the second order in the electron-radiation coupling. A complete description about how to derive it can be found in Blume.\cite{blume1,blume}
Neglecting relativistic effects, the interaction term is the usual ${\cal H}'\propto \sum_i {\bf p}_i \cdot {\bf {\cal A}}$, where ${\bf {\cal A}}$ is the vector potential for the electromagnetic field and ${\bf p}_i$ is the linear momentum of the $i^{th}$ atom.
The scattering amplitude close to a resonance is:

\begin{equation}
S_j({\bf q},{\bf k},\omega)= \sum_{mn} p_m \frac{\langle\Psi_m^{(j)}|{\hat O}^{\dag}({\bf \epsilon}',{\bf k}')|\Psi_n\rangle \langle\Psi_n|{\hat O}({\bf \epsilon},{\bf k})|\Psi_m^{(j)}\rangle}
 {\hbar\omega-(E_{nm})-i\Gamma}
\label{resonance}
\end{equation}

The operator ${\hat {O}}\equiv {\bf {\epsilon}} \cdot {\bf r} 
(1 + \frac{i}{2} {\bf k} \cdot {\bf r})$ is the usual matter-radiation interaction 
operator expanded up to the quadrupolar term, and ${\bf {\epsilon}}$ is the photon polarization. Primed quantities refer to outgoing objects.
 We neglect, as usual in the x-ray range, the magnetic dipole contribution. 
At thermal equilibrium at a temperature $T$, $p_m \equiv e^{-E_m/k_B T}/Z$ 
where $E_{m}$ is the energy of a state with wavefunction  $\Psi_m^{(i)}$,
 and $Z$ is the partition function.
The sum extends over all the excited states $\Psi_n^{(i)}$ with energy $E_n$. 
Finally, $E_{nm}\equiv E_n-E_m$, the index $i$ indicates the lattice 
site of the photoabsorbing atom and $\omega$ 
is the energy of the incoming photon.
In the dipole-dipole (E1-E1) approximation, Eq. (\ref{resonance}) becomes:

\begin{eqnarray}
S_{E1-E1}({\bf q},{\bf k},\omega)= \sum_{\alpha\beta} \epsilon'^*_{\alpha}\epsilon_{\beta} \nonumber \\
\sum_{mn} P_{mn} \langle\Psi_m|r_{\alpha}|\Psi_n\rangle \langle\Psi_n|r_{\beta}|\Psi_m\rangle
\label{e1e1}
\end{eqnarray}

\noindent where $P_{mn}\equiv p_m/(\hbar\omega-(E_{nm})-i\Gamma)$.

\noindent In the following we introduce the notation: $\langle\Psi_m|r_{\alpha}|\Psi_n\rangle \langle\Psi_n|r_{\beta}|\Psi_m\rangle \rightarrow D^{\alpha}_{mn}D^{\beta}_{nm}$.

We notice that the average in Eq. (\ref{e1e1}) can  be taken 
also on ${\overline{\Psi}}_m$ and ${\overline{\Psi}}_n$ which 
are time-reversed
with respect to $\Psi_n$(the corresponding energies are 
$E_{\overline{m} \overline{n}}$), provided we
 consider also the time-reversed distribution function $p_{\overline{m}}$.

Thus Eq. (\ref{e1e1}) can be re-written as:

\begin{eqnarray}
S_{E1-E1}({\bf q},{\bf k},\omega)= \frac{1}{2}\sum_{\alpha\beta} \epsilon'^*_{\alpha}\epsilon_{\beta} \sum_{mn} \nonumber \\
\big( P_{mn} D^{\alpha}_{mn}D^{\beta}_{nm} + P_{\overline{m} \overline{n}} D^{\beta}_{mn}D^{\alpha}_{nm})
\label{e1e1bis}
\end{eqnarray}

 Eq. (\ref{e1e1bis}) can be rewritten as the sum of a time-reversal even and a time-reversal odd part, $S^+_{E1-E1}$ and $S^-_{E1-E1}$, respectively, with:

\begin{equation}
S^{\pm}_{E1-E1}({\bf q},{\bf k},\omega)= \sum_{\alpha\beta} \epsilon'^*_{\alpha}\epsilon_{\beta} C^{\pm}_{\alpha\beta}(E1-E1)
\label{e1e1pm}
\end{equation}

and

\begin{eqnarray}
C^+_{\alpha\beta}(E1-E1) =\frac{1}{4} \sum_{mn} (P_{mn}+P_{\overline{m} \overline{n}}) (D^{\alpha}_{mn}D^{\beta}_{ nm} \nonumber \\
+D^{\beta}_{ mn}D^{\alpha}_{ nm}) = \frac{1}{2} \sum_{mn} (P_{mn}+P_{\overline{m} \overline{n}}) {\Re}(D^{\alpha}_{ mn}D^{\beta}_{ nm})
\label{e1e1plus}
\end{eqnarray}

\begin{eqnarray}
C^-_{\alpha\beta}(E1-E1) =\frac{1}{4} \sum_{mn} (P_{mn}-P_{\overline{m} \overline{n}}) (D^{\alpha}_{ mn}D^{\beta}_{ nm} \nonumber \\
-D^{\beta}_{ mn}D^{\alpha}_{ nm}) = \frac{i}{2} \sum_{mn} (P_{mn}-P_{\overline{m} \overline{n}}) {\Im}(D^{\alpha}_{ mn}D^{\beta}_{ nm})
\label{e1e1minus}
\end{eqnarray}
$\Re ,\Im$ stand for the real and imaginary parts
respectively of the arguments.

 When time-reversal is
a symmetry of the problem $E_{\overline{m} \overline{n}}= E_{mn}$ and
$ p_{\overline{m}}= p_m$. Therefore, under time-reversal $P_{mn}\rightarrow P_{\overline{m} \overline{n}}$.
Equation (\ref{e1e1minus}) shows that $C^-_{\alpha\beta}=0$ 
unless $P_{mn} \neq P_{\overline{m} \overline{n}}$. 
This latter condition is realized if the states 
$\Psi_m$ and ${\overline{\Psi}}_m$ are not degenerate, ie, if time-reversal is broken.

For the mixed dipole-quadrupole channel (E1-E2),  Eq. (\ref{resonance}) gives:

\begin{eqnarray}
S_{E1-E2}({\bf q},{\bf k},\omega)= \sum_{\alpha\beta} \epsilon'^*_{\alpha}\epsilon_{\beta} \sum_{mn} P_{mn} \nonumber \\
(ik_{\gamma} \langle\Psi_m|r_{\alpha}|\Psi_n\rangle \langle\Psi_n|r_{\beta}r_{\gamma}|\Psi_m\rangle \nonumber \\
-ik'_{\gamma}\langle\Psi_m|r_{\alpha}r_{\gamma}|\Psi_n\rangle \langle\Psi_n|r_{\beta}|\Psi_m\rangle) 
\label{e1e2}
\end{eqnarray}

Again, rewriting Eq. (\ref{e1e2}) in terms of the time-reversed states and 
taking the mean gives

\begin{eqnarray}
S_{E1-E2}({\bf q},{\bf k},\omega)= \frac{i}{2} \sum_{\alpha\beta} \epsilon'^*_{\alpha}\epsilon_{\beta} \\
\sum_{mn}  \big( P_{mn} (D^{\alpha}_{ mn}Q^{\beta\gamma}_{ nm} k_{\gamma} 
- Q^{\alpha\gamma}_{ mn}D^{\beta}_{ nm} k'_{\gamma}) \nonumber \\
+P_{\overline{m} \overline{n}} (Q^{\beta\gamma}_{ mn}D^{\alpha}_{ nm} k_{\gamma}-D^{\beta}_{ mn}Q^{\alpha\gamma}_{ nm} k'_{\gamma} ) \big) \nonumber
\label{e1e2bis}
\end{eqnarray}

\noindent where, we used the notation 
$\langle\Psi_m|r_{\alpha}|\Psi_n\rangle \langle\Psi_n|r_{\beta}r_{\gamma}|\Psi_m\rangle \rightarrow D^{\alpha}_{ mn}Q^{\beta\gamma}_{ nm}$.

It is possible to write a time-reversal even and a time-reversal odd
 contribution in the form:

\begin{eqnarray}
S^{\pm}_{E1-E2}({\bf q},{\bf k},\omega)= 
\sum_{\alpha\beta\gamma} \epsilon'^*_{\alpha}\epsilon_{\beta} \big( (k_{\gamma}+k'_{\gamma}) C^{\pm+}_{\alpha\beta\gamma}(E1-E2) \nonumber \\
 +(k_{\gamma}-k'_{\gamma}) C^{\pm-}_{\alpha\beta\gamma}(E1-E2)\big)
\label{e1e2pm}
\end{eqnarray}

where

\begin{eqnarray}
C^{++}_{\alpha\beta\gamma}(E1-E2) =
\frac{i}{2} \sum_{mn} (P_{mn}+P_{\overline{m} \overline{n}}) \nonumber \\
{\Re}(D^{\alpha}_{ mn}Q^{\beta\gamma}_{ nm}- 
D^{\beta}_{ mn}Q^{\alpha\gamma}_{ nm})
\label{e1e2plusplus}
\end{eqnarray}

\begin{eqnarray}
C^{+-}_{\alpha\beta\gamma}(E1-E2) =
\frac{i}{2} \sum_{mn} (P_{mn}+P_{\overline{m} \overline{n}}) \nonumber \\
\Re(D^{\alpha}_{ mn}Q^{\beta\gamma}_{ nm}+
D^{\beta}_{ mn}Q^{\alpha\gamma}_{ nm})
\label{e1e2plusminus}
\end{eqnarray}

and

\begin{eqnarray}
C^{-+}_{\alpha\beta\gamma}(E1-E2) =-\frac{1}{2} \sum_{mn}
 (P_{mn}-P_{\overline{m} \overline{n}})  \nonumber \\
{\Im}(D^{\alpha}_{ mn}Q^{\beta\gamma}_{ nm}+ 
D^{\beta}_{ mn}Q^{\alpha\gamma}_{ nm}) 
\label{e1e2minusplus}
\end{eqnarray}

\begin{eqnarray}
C^{--}_{\alpha\beta\gamma}(E1-E2) =-\frac{1}{2} \sum_{mn}
 (P_{mn}-P_{\overline{m} \overline{n}}) \nonumber \\
\Im(D^{\alpha}_{ mn}Q^{\beta\gamma}_{ nm}-
D^{\beta}_{ mn}Q^{\alpha\gamma}_{ nm})
\label{e1e2minusminus}
\end{eqnarray}

What is measured experimentally is of course the intensity, ie, 
\begin{eqnarray}
I({\bf q},{\bf k},\omega) = |F({\bf q},{\bf k},\omega)|^2.
\end{eqnarray}

To calculate $I({\bf q},{\bf k},\omega)$, one must first add the contributions $S_{{E1-E1}}$ and 
$S_{{E1-E2}}$ on the right side of Eq. (1) and then take the square of the 
absolute magnitude.
Note that the time-reversal even part of the scattering amplitude is real in the E1-E1 
channel and imaginary in the E1-E2 channel, 
while the opposite is true for the time-reversal odd parts. This means that in the
intensity for scattering of linearly polarized waves there is no interference term between
 the E1-E1 and E1-E2 channels when time-reversal is preserved but an interference exists
between the time-reversal even part of the E1-E1 and the time-reversal odd part of the
E1-E2 channel. 
This has important consequences for the detection of
 the magnetoelectric phase in cuprates materials, as we shall see later. 
(For circular polarizations an interference term exists between the time-reversal
even and the time-reversal odd parts in the E1-E1 channel alone.)

For the case of the cuprates no cartesian second-rank polar time-reversal breaking tensors
 exist for the
symmetries predicted state\cite{cmv1,simon} but third rank polar time-reversal breaking tensors
do exist (see the Appendix). Therefore one must rely on the time-reversal odd part of the E1-E2 term
and its interference with the usual E1-E1 (i.e. its time-reversal even part) for the detection
of the predicted states.
The polarization selection rules in the E1-E2 channels given 
by Eqs. (\ref{e1e2minusplus}), (\ref{e1e2minusminus}) are used in Sec. IV. 

The expected order
 of magnitude for the time-reversal breaking signal may be estimated as follows.
A general Time-reversal breaking wavefunction in a metal is of the
 form $\Psi_{\bf k}({\{r\}})$
with

\begin{equation}
\Psi_{\bf k}(\{r\}) = \Psi_{{\bf k}1}(\{r\}) + i\theta \Psi_{{\bf k}2}(\{r\})
\label{bloch}
\end{equation}

\noindent where $\Psi_{{\bf k}1}(\{r\}),\Psi_{{\bf k}2}(\{r\})$ may be taken to be pure
real in an appropriate gauge for the case that the {\it charge-density} has
inversion symmetry and where the complete wavefunction
$\Psi_{\bf k}(\{r\})$ cannot be transformed to a real function by any
gauge transformation.
In order  to obtain a finite time-reversal odd part, $C^-$, the product of the
two matrix element in Eq. (\ref{resonance}) must have an imaginary part.\cite{luo,blume1}
 Thus, either the
first or the second matrix element in the product must be
imaginary and since the core wavefunction is a real function,
it then follows that the time-reversal odd scattering amplitude is 
of order $\theta$ with respect
 to the corresponding time-reversal even part.
In the case the signal to be detected is in the E1-E2 channel, 
there is an extra factor of about 0.1-0.2 with respect to the E1-E1 main edge,
 due to the relative weight of the radial density of states.
From Refs. [\onlinecite{cmv2,simon}] one can get a rough estimate for 
$\theta \simeq 0.05$.
To get an interference term  
 $\simeq \theta$, one needs to have
  an interference with the main E1-E1 term,
 which is real (its coefficient, $\epsilon'^* \cdot \epsilon$
 is real for both circular and linear polarizations). Since the quadrupole matrix elements are
are usually of $O(10^{-1})$ of the dipole matrix elements, the magnitude of the
effect is of $O(5\times 10^{-3})$ of the typical resonant Bragg peaks.

\section{Time-reversal breaking processes in linear and circular dichroism.}

The atomic x-ray absorption cross section can be written as the imaginary part of the forward scattering amplitude or by the Fermi golden rule:

\begin{eqnarray}
\sigma_i=4\pi^2 \alpha \hbar \omega \sum_{mn} p_m \langle \Psi_m ^{(i)}|
{ \hat {O}}^{\dag} | \Psi_n^{(i)} \rangle \langle \Psi_n ^{(i)}|
{ \hat {O}} | \Psi_m^{(i)} \rangle \nonumber \\
 \delta(\hbar\omega-(E_n-E_m))
\label{absor}
\end{eqnarray}

\noindent where $\alpha$ is the fine-structure 
constant.
The total absorption cross-section is then given by the sum over all the
atoms in the unit cell.
In the following two subsections we describe two dichroism experiments 
 to detect the time-reversal broken phase in
the pseudogap region of the cuprates. The geometrical conditions for either are
determined by the magnetic symmetries of the material as well as the experimental
 geometry. The intensity of the effect
is then simply determined on the basis of the so-called ``sum rules''
\cite{carra,natoli1,natoli2} and of the explicit form of the
conduction-band wave function, as determined for the cuprates.\cite{cmv1,simon}

\subsection{Nonreciprocal linear dichroism}

Linear dichroism in the x-ray range is the differential spectrum obtained 
by subtracting two core absorption spectra taken with orthogonal linear 
polarizations. If the difference is of dipole-dipole (E1-E1) origin, it 
 represents normal X-ray Linear Dichroism (XLD) and if of dipole-quadrupole origin,
it represents 
 NonReciprocal Linear Dichroism (NRXLD). 

From Ref.[\onlinecite{natoli1,natoli2}]  the sum rules for
XLD and NRXLD at the K-edge are:

\begin{eqnarray}
\sigma_{XLD} \propto  <\Psi^{(i)}_{E}|(L^2_{x'}-L^2_{y'})|\Psi^{(i)}_{E}>
\label{lindice1e1}
\end{eqnarray}
\begin{eqnarray}
\sigma_{NRXLD} \propto  <\Psi^{(i)}_{E}|(L^2_{x'}-L^2_{y'})\Omega_z'|\Psi^{(i)}_{E}> - c.c.
\label{lindice1e2}
\end{eqnarray}

\noindent where $x'$ and $y'$ are the two arbitrary orthogonal directions of
the two polarizations and $z'$ is the direction of the photon wavevector
$\bf{k}$. $\bf{L}$ is the usual angular momentum operator and the anapole
operator $\bf{\Omega} \equiv (\bf{L} \wedge \bf{n} - \bf{n} \wedge
\bf{L})$, where $\bf{n}$ is the unit radial vector operator.\cite{carra,varsha}
The averages of these operators are taken integrating over all the scattering (excited) states
corresponding to the energy $E$ at which the spectrum is recorded.

The main difference between Eq. (\ref{lindice1e1}) and (\ref{lindice1e2})
is in the time-reversal and inversion properties of the two operators.
The first operator is time-reversal even and inversion even, being the
symmetric traceless part of a cartesian second rank polar tensor. On the
other hand the second operator is time-reversal odd and inversion odd: its mean value over the states $\Psi^{(i)}(E)$ is different from
zero only if both these symmetries are broken, ie, if the system under
consideration is {\it magnetoelectric}. This situation is fulfilled in 
one of the time-reversal broken phases found in Refs. [\onlinecite{cmv1,simon}]. Thus the detection of 
NRXLD would be a clear evidence of the time-reversal broken phase.
The geometrical conditions to detect the effect are discussed in detail in the next section.
The order of magnitude of the effect is readily estimated as in the previous section for the scattering amplitude.
The NRXLD is due to an electric-dipole electric-quadrupole transition. As
such, it mixes states of different parity: considering a K-edge process, these states are the $p$ and $d$
conduction bands.
As shown in Refs. [\onlinecite{cmv1,simon}], the generic time-reversal broken wave function
for the cuprates is such that $\Psi_{\bf{k}2}(\{r\})$ is of $p$ type and in
$\Psi_{\bf{k}1}(\{r\})$ there is a part with $d$ symmetry. Thus the average in Eq. (\ref{lindice1e2}) is proportional to $\theta$
and the
ratio of the total dichroic signal with respect to the absorption edge is
proportional to $\theta$ times the ratio of radial density of states 
in the E1-E2 
 to the E1-E1 channels. 
This of course gives the same estimate obtained for the scattering amplitude,
 about 0.5$\%$,
which is detectable with modern synchrotron radiation sources.\cite{goulon}

\subsection{Magnetochiral dichroism}

Another technique that can be adopted to detect the time-reversal broken
phase in the cuprates is the MagnetoChiral Dichroism (MCD). Circular dichroism
in E1-E1 channel is excluded, being sensitive to the expectation value of the orbital angular momentum, which is zero for the patterns found in Refs. [\onlinecite{cmv1,simon}]. 
Inversion-odd circular dichroism
is time-reversal even, giving rise to x-ray natural circular dichroism, and thus it is excluded as well. But, as shown recently by
Goulon et al.\cite{goulon2} for Cr$_2$O$_3$, the sum of two right and left
circularly polarized spectra can give rise to a dichroism in magnetochiral
systems, ie, magnetoelectric systems whose magnetoelectric tensor has an
antisymmetric off-diagonal part.
Note that this is just the case of the magnetic symmetry predicted by Simon
and Varma.\cite{simon}
In this case the dichroism is defined as the difference between the sum of
two orthogonally polarized spectra in one magnetoelectric
domain and those in the opposite magnetoelectric domain.
Note that for cuprates the choice of the magnetoelectric domain could be made difficult by the fact that they are not insulators.

As shown in Ref. [\onlinecite{dimatteo}], the magnetochiral dichroism is
proportional to the expectation value, in the conduction states, of the
operator $L^2_{z'}\Omega_{z'}$, where $z'$ is the direction of the wavevector of
the absorbed photon:

\begin{eqnarray}
\sigma_{MCD} \propto  <\Psi^{(i)}_{E}|L^2_{z'}\Omega_{z'}|\Psi^{(i)}_{E}>
\label{cirdice1e1}
\end{eqnarray}

In section IV we analyze the geometrical conditions to detect the effect. An eventual magnetochiral signal
is a definite indication of the time-reversal broken phase.

As for NRXLD, the intensity of the signal should be of order 0.5$\%$,
 within the present experimental capabilities.

\begin{figure}
\centerline{\epsfig{file=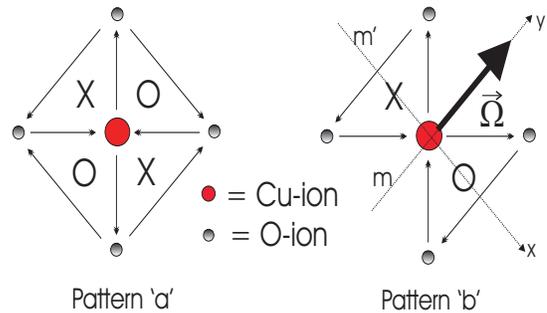,height=1.6in,width=2.8in}}
\vspace{5pt}
\caption
{The two time-reversal breaking orbital-current patterns. For pattern 'b', magnetoelectric, the allowed direction of the orbital anapole is shown.}
\label{cell}
\end{figure}

\section{Application to the Cuprates}

We have considered the following materials: HgBa$_2$CuO$_{4+\delta}$,
 Tl$_2$Ba$_2$CuO$_{6+\delta}$, La$_{1-\delta}$Sr$_{\delta}$CuO$_4$, Bi$_2$Sr$_2$CaCu$_2$O$_8$. Except the orthorombic phase of  La$_{1-\delta}$Sr$_{\delta}$CuO$_4$, that must be studied separately,
these cuprates all have {\it crystalline} space group symmetry
4/mmm (the crystal class is primitive for
 HgBa$_2$CuO$_{4+\delta}$, body-centered for all the others).
 The copper ion occupies a site of maximal symmetry (4/mmm) 
in all the cases except for  Bi$_2$Sr$_2$CaCu$_2$O$_8$, 
where the symmetry is 4mm: the mirror plane orthogonal to
 the four-fold axis is lost because in the bilayered compounds
 one of the apical oxygens is missing. Two types of time-reversal breaking states can be
proposed, according to the results of Ref. [\onlinecite{simon}], one with magnetic-group symmetry ${\underbar 4}/{\underbar m}$mm and
the other with magnetic-
 group symmetry m${\underbar m}$m (we use Birss' notation ${\underbar O}$ for a time-reversed symmetry operator $O$). 

The magnetic group 
${\underbar 4}/{\underbar m}$mm, (which remains tetragonal) 
 is such that the 
inversion symmetry is preserved, while 
both time-reversal and the product of 
the time-reversal and inversion are broken. 
Since inversion is preserved, 
no E1-E2 signals (that are inversion-odd) are possible.
 At the same time, the simultaneous  presence of three 
mirror planes as symmetry elements (even if one is time-reversed)
forbids the possibility of a magnetic moment as a local invariant 
(because the magnetic moment is an axial quantity and cannot be
 invariant under the action of two orthogonal mirror planes). 
This implies that with this pattern time-reversal odd quantities can be detected only in the E2-E2 
channel.

More interesting is the case when the symmetry is 
reduced to m${\underbar m}$m. This is the symmetry consistent with the 
results of the ARPES experiment in Bi-2212. First of all, 
in this case the system is no longer tetragonal. This can in principle be 
detected by an azimuthal scan around the original four-fold axis as 
the new periodicity is determined by the mirror and is no longer 
four-fold. Since inversion symmetry is also lost, 
in this phase  the detection of some phenomena linked to 
magnetoelectricity becomes possible. Apart from a direct observation 
of magnetoelectric susceptibility, x-ray 
diffraction, non-reciprocal 
linear dichroism and magnetochirality in the E1-E2 
channel are all possible experiments to observe this phase.

As we already saw the interesting effects arise in the E1-E2 channel through  finite
third-rank polar time-reversal odd tensors specified, for example, in
Eqs. (\ref{e1e2minusplus}), (\ref{e1e2minusminus}). We specify the tensors using the following co-ordinates:
      The c axis of the crystal coincides with the $\hat{z}$ direction, a and b crystalline axis (those along the nearest
neighbor Cu-Cu vector)
      are along the $(\hat{x}+\hat{y})/\surd 2$ and $(\hat{x}-\hat{y})/\surd 2$  directions,
as is clear from Fig. 1b. The point group m${\underbar m}$m has the following elements:
the identity,  two-fold axis rotation around  $\hat{y}$ ($C_y$), the reflections 
in the $x-y$ and $y-z$ planes ($\sigma_z$ and $\sigma_x$), $RI$, $R\sigma_y$, $R C_x$ 
and $R C_z$ (where $R$ is the time-reversal operator). The distinct non-zero elements of the third-rank polar tensors, 
odd under time-reversal in this point group are seven\cite{birss}: $C^-_{yyy},~C^-_{xxy},~C^-_{zzy},~C^-_{xyx},~C^{-}_{yxx},~C^{-}_{zyz},~C^{-}_{yzz}$.

If we write the seven allowed tensors by symmetrizing the first two indices, we get the symmetric 
$C^-_{yyy}$,  $C^-_{xxy}$, $C^-_{zzy}$,  $C^-_{xyx}+C^-_{yxx}$ and  $C^-_{zyz}+C^-_{yzz}$ and the antisymmetric  $C^-_{xyx}-C^-_{yxx}$ and  $C^-_{zyz}-C^-_{yzz}$. The formers can be detected only with the symmetrized part of the polarizations, ie, they correspond to $C^{-,+}$ (Eqs. (\ref{e1e2minusplus}) and (\ref{e1e2minusminus})) and they are proportional to ${\bf k} + {\bf k'}$. They also correspond to tensors ${\tilde {A}}^{(1)}$ and $ {\tilde {C}}^{(k)}$ in the Appendix and as such they can be detected in both $\sigma\sigma$ and $\sigma\pi$ polarizations (${\tilde {A}}^{(1)}$  only in $\sigma\sigma$).
The latters correspond to the tensor ${\tilde {B}}^{(2)}$ in the Appendix. Even if proportional to  ${\bf k} - {\bf k'}$, the signal is not maximized in a backscattering configuration, where it would be zero, because if ${\bf k}$ and ${\bf k'}$ are both parallel to $z$, then $\epsilon_z=0$. This signal can be detected only in the $\sigma\pi$ channel.
Thus, for the pattern depicted in Fig. 1b, the preferred geometrical configuration should be when the wave vectors lie in the $x-z$ plane, with  ${\bf k} - {\bf k'}$ along $z$ and  ${\bf k} + {\bf k'}$ along $x$, both about 45$^o$ from the two axes, in order to optimize the projections of ${\bf k}$ and of the outgoing polarization on the $z$-axis. The incoming polarization in the $\sigma\pi$ channel is along $y$ and the outgoing polarization is in the $x-z$ plane, orthogonal to ${\bf k'}$. In this situation the only contributions come from the antisymmetric part of $C^-_{yzz}$ and from the symmetric part of $C^-_{yxx}$. An azimuthal scan about the $z$-axis could reveal the two-fold time-reversal breaking signal with respect to the four-fold time-reversal even one.
These considerations are, of course, subject to the requirement that ${\bf k}-{\bf k'}={\bf q}$, where ${\bf q}$ is a Bragg vector.

Another possibility to detect the time-reversal breaking signal is to exploit the different polarization dependence of $C^{--}_{\alpha\beta\gamma}(E1-E2)$ and $C^+_{\alpha\beta}(E1-E1)$. The former is antisymmetric in the exchange of the two polarizations, while the latter is symmetric. 
Also the E1-E1 contribution antisymmetric in the polarizations (ie, $C^-_{\alpha\beta}(E1-E1)$) is zero, because it is proportional to the average magnetic moment on the Cu site, while the time-reversal even E1-E2 term, antisymmetric in the polarizations (ie, $C^{++}_{\alpha\beta\gamma}(E1-E2)$) is imaginary.
Therefore, subtracting the signals obtained with switched orthogonal incoming and outgoing polarizations leaves only $2 C^+_{\alpha\beta}(E1-E1) C^{--}_{\alpha\beta\gamma}(E1-E2)$. Observing the effect requires a sensitivity better than $10^{-3}$.

The orthorombic phase of  La$_{1-\delta}$Sr$_{\delta}$CuO$_4$
 has different symmetries, but the conclusions are similar, 
as well as the experimental techniques to detect the 
time-reversal breaking. In the more symmetric phase, 
due to the buckling of the oxygens, the symmetry 
around each copper atom is 2/m (and the space crystal symmetry is Cmce).
 This symmetry is lowered, by the magnetic orbital patterns to 
${\underbar 2}/ {\underbar m}$ or to one of the two magnetoelectric
 ${\underbar 2}$/ m, $2/ {\underbar m}$. For the first case we can 
apply, even with a lower symmetry, the same considerations as for 
the ${\underbar 4}$/mm ${\underbar m}$, while for the two 
magnetoelectric phases, being also magnetochiral, 
the same considerations as for m${\underbar m}$m apply.

We have already discussed the considerations regarding the magnitudes of the
intensity of order $\theta$ due to time-reversal breaking.
There is of course a contribution of $O(\theta^2)$ also. 
Apart from other considerations of domains noted below, time-reversed domains 
in the scattering volume are expected to affect adversely the effect of $O(\theta)$
and not of $O(\theta^2)$.
Besides the estimate of the intensity of the signal, 
it is important to analyze how the geometry of the systems can 
influence the choice of the experimental technique. The origin of the 
time-reversal breaking is due to  orbital currents circulating around 
each copper ion with a well-defined pattern 
 in each CuO$_2$ plane, as described in Ref. [\onlinecite{cmv1,simon}]. 
The ARPES experiments probe only the surface Cu-O layers and 
indicate a surprisingly large domains size, greater than about 10 microns.
If we accept this as correct, the next question concerns the relative order
 from plane to plane. The inter-plane coupling is expected to be quite weak and
 answer to this question cannot be reliably obtianed from theoretical considerations.
We shall examine the case of both ``ferromagnetic'' (FM) and "antiferromagnetic" (AFM) ordering
between the planes.

For all the cases we have examined, the ferromagnetic pattern preserves the
size of the primitive cell and the antiferromagnetic pattern changes it. Therefore 
the "forbidden" diffraction spots (at h+k+l = odd for body-centered tetragonal or
 body centered orthogonal lattices) stay "forbidden'" for 
the FM pattern but become allowed for
 the AFM patterns. If the ordering between the planes is random, as
 is not unlikely, Bragg-rods are expected to arise.
For the AFM pattern dichroic techniques are useless,
 as the signal averages to zero in the unit cell. 
In this case only x-ray scattering can detect a signal, 
because the ordinarily forbidden Bragg-spots  can be revealed 
by means of linearly polarized light in the E1-E2 channel (for pattern b of Fig. 1).
 In case the alignment of successive layers is purely ferromagnetic,
it is rather complicated to resolve the magnetic signal
 with respect to the non-magnetic by x-ray scattering, because of
 smallness of the former ($10^{-3}$).
 One possible way to perform the experiment in this case
 is linked to the different azimuthal scan between the normal
 (4-fold periodic) and the magnetoelectric (two-fold periodic)
 phases, as shown in the following subsection.
Probably more appropriate to the case of purely FM ordering is the use of dichroism.

In the next subsection we shall deal with the specific case of Bi-2212, that is specially note worthy in this respect,   since 
 the evidence for the symmetry breaking phase has first been 
 discovered in it.\cite{kaminski}

\subsection{The case of Bi$_2$Sr$_2$CaCu$_2$O$_8$.}

The space group given is the tetragonal I4/mmm with 
two formula units in
the cell. The four Cu ions are on the  Wyckoff sites
4e, whose symmetry is  4mm, ie, not the full space symmetry
of the cell (full symmetry that is recovered only by the Ca ions, on the 2a
sites). The reason for this is that the Cu ions in a bilayered
system like Bi-2212 do not have one of the two apical oxygens and thus the /m plane,
that is orthogonal to the 4-fold axis is missing. For this reason they have
not even a centre of inversion and the situation is quite different from
the mono-layered compound, like  HgBa$_2$CuO$_{4+\delta}$, where the 
inversion and the /m plane are present.

From the experiment of
 Kaminski {\it et al.}\cite{kaminski}, we can only conclude that the bilayers on the surface cells
have currents in phase (in the examined region) but nothing can be inferred for
 phases between different planes in the bulk region.

If we superimpose one of the two current structures, the {\it local}
 magnetic symmetry on the Cu ion is reduced
from 4mm$\otimes {R}$  to ${\underline {4}}$mm
 (still tetragonal)  for pattern $a$ of Fig. 1, while it
 is reduced to
 ${\underline {2}}$m${\underline {m}}$ 
for pattern $b$. In this latter case the
magnetic symmetry is orthorhombic (the four-fold axis is lost)
 and the
mirror planes are no longer orthogonal to the Cu-O bonds, but to $x$ and $y$ of Fig. 1b..
We already dealt with the AFM case: let us now examine the case when all CuO$_2$ layers are
 in phase (FM ordering).
If all the four Cu ions have a current pattern of type $a$,
 the space magnetic symmetry is reduced to  ${\underline {4}}/{\underline {m}}$mm: the group
is still tetragonal and, even if Cu sites are not centrosymmetric, there is a global inversion centre 
in correspondance
to the Ca ions. Because of the global centre of inversion, 
it is not possible to detect E1-E2 terms in an absorption 
experiment (their average gives zero). The only
signal is in the E2-E2 channel and its 
intensity is expected to be very small (there is an extra factor due to the
 radial matrix elements of the E2-E2 channel with respect
 of the E1-E2, that can be estimated as an extra 0.1).

If all the four Cu ions have a current pattern of type 
$b$, the space magnetic symmetry is reduced to m${\underline {m}}$m, as already seen. 
This group does not
contain the time-reversal or the inversion as separated 
symmetries, but it contains
their product. It is a magnetoelectric group (contrary to the original
4/mmm) and, as the corresponding magnetoelectric tensor has an antisymmetric
off-diagonal part, it admits also magnetochirality. 
For this reason this configuration is even more interesting in that
 it can be revealed by linear (non-reciprocal) and magnetochiral 
dichroism. 

Here we determine the best geometrical conditions for such experiments.
We refer to Fig. 1b and to Eqs. (\ref{lindice1e2}), (\ref{cirdice1e1}).
It is clear that for both kind of dichroisms the signal is maximized when the light is shone in the same direction as the anapole, as in both Eqs. (\ref{lindice1e2}), (\ref{cirdice1e1}) $\Omega_{z'}$ stands for ${\bf \Omega}\cdot {\bf k}$. The anapole, being a time-reversal odd polar vector, is invariant under a parallel mirror symmetry and changes sign if the mirror is orthogonal. The contrary happens if the mirror is also time-reversed. It follows that the only invariant direction for such a vector is the one drawn in Fig. 1b, orthogonal to the time-reversed mirror plane.
Thus, in order to have a non-zero time-reversal odd signal, the light must have a wave vector with a component along this direction.
Notice that, in this case, a linear dichroism of E1-E1 origin is present, because of the anisotropy of two orthogonal directions within the time-reversed mirror plane, and it is expected to be bigger than the time-reversal breaking one. In order to detect the latter, two experiments must be performed, with opposite magnetoelectric domains: the substraction of the two spectra gives then twice the E1-E2 NRXLD, being the E1-E1 XLD invariant in the two domains.
Similar considerations also hold for the magnetochiral dichroism, ie, in order to maximize the signal, the light must be shone in the direction of the anapole: an useful description of the experiment is then given in Ref. [\onlinecite{goulon2}].

Finally, the reduction of the group from tetragonal to orthorombic,
 with the pattern $b$, allows, in principle, the detection of the signal
 also by means of x-ray resonant scattering. The trick, as sketched before, is the 
following: even if all the allowed magnetic reflections
 correspond to non-magnetic allowed reflections, too, these 
latter have a four-fold symmetry around the axis orthogonal to the CuO$_2$ plane. Consider 
a reflection, like the (002), whose exchange vector is directed
orthogonal to the CuO$_2$ planes. In the normal phase, making 
an azimuthal scan around this vector one obtains a four-fold 
periodicity, ie, four minima.
The symmetry reduction due to the magnetic pattern (two-fold) reduces the number of 
minima to two. 
Thus deviations from a perfect four-fold azimuthal signal would 
be an indication of the magnetic effect. This effect is however quite difficult to
 measure, because the deviation is again expected to be of the order $5 \cdot 10^{-3}$.

\section*{Acknowledgments.}
CMV wishes to thank D.E. Moncton and G.Sawatzky for asking the question of whether
effects of time-reversal breaking in absorption experiments also have observable consequences
in x-ray diffraction experiments, to P.Carra for
 introducing him to the literature on x-ray dichroism and 
to M.E. Simon for useful discussions. 

SDM would like to thank C.R. Natoli and I. Marri for useful old discussions on x-ray scattering and dichroism

\section*{Appendix}

As in the text we dealt with both cartesian tensors and irreducible representations of these latters, it can be useful to list the way to express the ones in terms of the others. For the irreducible representations we shall adopt spherical coordinates, that turn out to be more versatile in this respect.

For the E1-E1 channel it is quite straightforward: the decomposition of a cartesian rank-two tensor into the irreducible spherical representations can be found in many textbooks.\cite{varsha}
Any rank-two polar tensor $D_{\alpha\beta}$ can be expressed in terms of its SO(3) irreducible representations:

1) a scalar, corresponding to the trace of  $D_{\alpha\beta}$: $S = D_{xx} + D_{yy} +D_{zz}$

2) a vector, that is the dual of the antisymmetric part of  $D_{\alpha\beta}$: $V_i = \epsilon_{ijk} D_{jk}$, where $\epsilon_{ijk}$ is the third-rank totally antisymmetric Levi-Civita tensor: it is proportional, at the K-edge, to the orbital angular momentum. For example, referring to Eq. (7), $L_z \propto C^-_{xy}-C^-_{yx}$

3) a symmetric traceless part: $T_{ab} = \frac{1}{2} (D_{ab}+D_{ba} -\frac{2}{3}\delta_{ab}(D_{xx}+D_{yy}+D_{zz}))$

Things turn out to be slightly more complicated when we deal with the E1-E2 channel. In this case there are $3^3 \times 2$, ie, 54 independent components (the factor two comes out from the fact that in the E1-E2 channel also the complex conjugate matrix elements are present). The explicit form of Eq. (8) shows that the tensor $Q^{\alpha\gamma}_{mn}$ must be symmetric, thus reducing the numebr of independent components to $18 \times 2$. They further reduce to $15 \times 2$ because the scalar coupling of $\alpha$ and $\gamma$ in  $Q^{\alpha\gamma}_{mn}$ cannot be detected as it is coupled to the scalar product ${\bf \epsilon} \cdot {\bf k} =0$.
In order to classify these components properly, it is much more convenient to classify the tensors representing the properties of the x-ray beam, that are coupled scalarly to the ones representing the properties of the matter.\cite{luo,carra2} This classification turns out to be material-independent, in that it only shows the tensorial propoerties to which the x-ray beam is sensitive in the E1-E2 channel. It must then be combined with the explicit symmetries of the material, as done in section IV, to get more useful informations.

The only allowed tensors, classified in terms of their time-reversal properties, are the following:

\begin{center}
$\begin{array}{|c||c|c|c||} \hline
  &  \epsilon'^* \cdot \epsilon \stackrel{\scriptsize{(+)}}{~~~~~} &  \epsilon'^* \wedge \epsilon \stackrel{(-)}{~~~~~} & [  \epsilon'^*, \epsilon ]^{(2)} \stackrel{(+)}{~~~~~} \\ \hline \hline 
{\bf k} + {\bf k'} \stackrel{(-)}{~~~~~} & {\tilde {A}}^{(1)} &  B^{(2)} &  {\tilde {C}}^{(1)}, ~ {\tilde {C}}^{(2)}, ~ {\tilde {C}}^{(3)} \\ \hline 
{\bf k} - {\bf k'} \stackrel{(+)}{~~~~~} & A^{(1)} & {\tilde {B}}^{(2)} & C^{(1)}, ~ C^{(2)}, ~  C^{(3)} \\ \hline
\end{array}$
\end{center}

The superscripts (+) and (-) refer to the time-reversal behaviour of the corresponding quantities and $[\epsilon'^*, \epsilon ]^{(2)}$ is the $\epsilon'$-$\epsilon$ coupling to a rank-two irreducible tensor. Non-tilded tensors are time-reversal even and tilded tensors are time-reversal odd.
Their explicit expressions in spherical coordinates is (apart from multiplicative prefactors):

$$ {\tilde {A}}^{(1)}_q =  \epsilon'^* \cdot \epsilon (k_q + k'_q)$$
$$ A^{(1)}_q =  \epsilon'^* \cdot \epsilon (k_q - k'_q)$$
$$ B^{(2)}_q = \sum_{mm'} C^{112}_{mm'q} (\epsilon'^* \wedge \epsilon)_m (k_{m'} + k'_{m'})$$
$$ {\tilde {B}}^{(2)}_q = \sum_{mm'} C^{112}_{mm'q} (\epsilon'^* \wedge \epsilon)_m (k_{m'} - k'_{m'})$$
$$ {\tilde {C}}^{(k)}_q = \sum_{mm'} C^{21k}_{mm'q} [\epsilon'^*, \epsilon]^{(2)}_m (k_{m'} + k'_{m'})$$
$$ C^{(k)}_q = \sum_{mm'} C^{21k}_{mm'q} [\epsilon'^*, \epsilon]^{(2)}_m (k_{m'} - k'_{m'})$$

\noindent where $ C^{ll'k}_{mm'q}$ are the Clebsch-Gordan coefficient of the coupling.
The corresponding expression in cartesian coordinates can be easily derived.\cite{varsha}

We just would like to note that, for example, one of the tensors, inversion and time-reversal odd, that can be detected in the cuprates, $C^{--}_{yzz}$, is coupled to $ \frac{1}{\sqrt{2}} ({\tilde {B}}^{(2)}_1 - {\tilde {B}}^{(2)}_{-1}) $: it corresponds to what  Carra {\it et al.}\cite{carra3} identified as an indication of the tensorial magnetic and electric displacement: $C^{--}_{yzz} \propto [{\bf L}, {\bf n}]^{(2)}_1 - [{\bf L}, {\bf n}]^{(2)}_{-1} $.

In the same way, the anapole ${\bf \Omega}$ is the vector coupled to ${\tilde {A}}^{(1)}$, time-reversal odd and symmetric in the polarization exchange: $\Omega_i = \sum_j C^{-+}_{jji}$.

\end{multicols}

\begin{references}

\bibitem{cmv1}
C.M. Varma,  Phys. Rev. B {\bf 55}, 14554 (1997); Physical Rev. Letters
{\bf 83}, 3538 (1999).
\bibitem{cmv2}
C.M.Varma, Phys. Rev. B {\bf 61}, R3804 (2000)
\bibitem{simon}
M.E. Simon and C.M. Varma, Phys. Rev. Lett., accepted for publication (2002); cond-mat/0201036
\bibitem{kaminski}
A. Kaminski, S. Rosenkranz, H. M. Fretwell, J. C. Campuzano, 
Z. Li, H. Raffy, W. G. Cullen, H. You, C. G. Olson, C. M. Varma, H. H\"ochst,
 Nature {\bf 416}, 610 (2002); cond-mat/0203133
\bibitem{LL}
L.D. Landau, E.M. Lifshitz and L. P. Pitaevskii, {\it Electrodyanmics of Continuous Media}, 
2nd edition, Butterworth-Heinemann, Oxford (1984).
\bibitem{birss}
R.R. Birss, {\it Symmetry and Magnetism}, edited by E.P. Wohlfarth (North-Holland, Amsterdam, 1964), p. 137, 104 and 60.
\bibitem{mes}
M.E. Simon and C.M.Varma, cond-mat/0210672.
\bibitem{carra1}
P. Carra, B.T. Thole, M. Altarelli, X. Wang, Phys. Rev. Lett. {\bf 70}, 694 (1993)
\bibitem{luo}
J. Luo, G.T. Trammell, J.P. Hannon,  Phys. Rev. Lett. {\bf 71}, 287 (1993)
\bibitem{trammel}
J.P. Hannon , G.T. Trammell, M. Blume and Doon Gibbs, Phys. Rev. Lett. {\bf 61}, 1245 (1988)
\bibitem{blume1}
M. Blume, {\it Magnetic effect in anomalous dispersion}, in ``Resonant anomalous x-ray scattering, edited by G. Materlik, C.J. Sparks and K. Fischer (Elsevier Science B.V., 1994), p. 495
\bibitem{blume}
M. Blume, J. Appl. Phys. {\bf 57} (1985) 3615
\bibitem{carra}
P. Carra, R. Benoist, Phys. Rev. B {\bf 62}, R7703 (2000)
\bibitem{natoli1}
S. Di Matteo, C.R.Natoli,  J. Synchr. Rad. {\bf 9}, 9 (2002)
\bibitem{natoli2}
S. Di Matteo, C.R. Natoli, submitted to Phys. Rev. B (2002)
\bibitem{varsha}
D.A. Varshalovich, A.N. Moskalev, V.K. Kersonskii, {\it Quantum Theory of Angular Momentum}, World Scientific (1988)
\bibitem{goulon}
J. Goulon, A. Rogalev, C. Goulon-Ginet, G. Benayoun, L. Paolasini, C. Brouder, C. Malgrange and P.A. Metcalf, Phys. Rev. Lett {\bf 85}, 4385 (2000)
\bibitem{goulon2}
J. Goulon, A. Rogalev, F. Wilhelm, C. Goulon-Ginet, P. Carra, D. Cabaret, C. Brouder, Phys. Rev. Lett. {\bf 88}, 237401 (2002)
\bibitem{dimatteo}
S. Di Matteo, C.R. Natoli, accepted for publication in Phys. Rev. B (2002)
\bibitem{carra2}
P. Carra, T. Thole, Rev. Mod. Phys. {\bf 66}, 1509, (1994)
\bibitem{carra3}
P. Carra, A. Jerez, I. Marri, pre-print: cond-mat/0210152


\end{references}
\end{document}